\documentclass[submission,copyright]{eptcs}

\providecommand{\event}{TTC 2013} % Name of the event you are submitting to
\usepackage{breakurl}             % Not needed if you use pdflatex only.

\usepackage[utf8]{inputenc}
\usepackage[T1]{fontenc}
\usepackage{url}
\usepackage{float}
\usepackage{wrapfig}
\usepackage{tabularx}
\usepackage{graphicx}
\usepackage{caption}
\usepackage{subcaption}
\graphicspath{ {./fig/} }

\usepackage{color}
\definecolor{listinggray}{rgb}{0.9,0.9,0.9}
\definecolor{keywordcolor}{rgb}{0.5,0,0.1}
\definecolor{commentcolor}{rgb}{0,0.3,0.1}
\definecolor{stringcolor}{rgb}{0,0,1}
\usepackage{listings}
\lstdefinelanguage{viatra}
{morekeywords={@QueryBasedFeature,@Constraint,count,pattern,
    neg,find,import,true,false,or,check,job,action,state,severity,location,message,
    oclIsKindOf,self,exists,includes,invariant,class},
 sensitive=true, morecomment=[l]{//}, morecomment=[s]{/*}{*/},
 morestring=[b]{"},
}
\newcommand{\listingIQPL}[1]
{
\lstset{basicstyle=\tiny\ttfamily}
\lstset{commentstyle=\itshape\color{commentcolor}\ttfamily}
\lstset{stringstyle=\color{stringcolor}\ttfamily}
\lstset{frameround=tttt}
\lstset{captionpos=b}
\lstset{keywordstyle=\color{keywordcolor}\bfseries\ttfamily}
\lstset{showstringspaces=false}
\lstset{tabsize=2}
\lstset{numbers=left,numberstyle=\tiny\ttfamily,stepnumber=1,numbersep=5pt}
\lstset{language=viatra}
\lstset{escapeinside={(*@}{@*)}}
\lstinputlisting{#1}
}

\lstdefinelanguage{Xtend}{
 morekeywords={def,val,var,cached,case,default,extension,false,import,JAVA,WORKFLOWSLOT,let,new,null,private,create,switch,this,true,reexport,around,if,then,else,context,DEFAULT_NO_UPDATE_AND_DISAPPEAR,DEFAULT,APPEARED,DISAPPEARED,UPDATED,processor},
 keywordstyle=[2]{\textbf},
 morecomment=[l]{//}, 
 morecomment=[s]{/*}{*/}, 
 morestring=[b]",
 tabsize=2
}
\newcommand{\listingXtend}[1]{
\lstset{numbers=left,numberstyle=\tiny\ttfamily,stepnumber=1,numbersep=5pt}
\lstset{commentstyle=\itshape\color{commentcolor}\ttfamily}
\lstset{stringstyle=\color{stringcolor}\ttfamily}
\lstset{keywordstyle=\color{keywordcolor}\bfseries\ttfamily}
\lstset{breaklines=true}
\lstset{language=Xtend}
\lstset{mathescape=true}
\lstset{basicstyle=\tiny\ttfamily}
\lstset{escapeinside={(*@}{@*)}}
\lstinputlisting{#1}
}

\lstdefinelanguage{Pseudo}{
 morekeywords={FOR, CALL, ENDFOR, IF THEN, ELSE, ENDIF},
 keywordstyle=[2]{\textbf},
 morecomment=[l]{//}, 
 morecomment=[s]{/*}{*/}, 
 morestring=[b]",
 tabsize=2
}

\definecolor{JavaKeyword}{rgb}{0.5,0,0.33}
\definecolor{JavaString}{rgb}{0.16,0,1}
\floatstyle{ruled}
\newfloat{listing}{htbp}{lop}
\floatname{listing}{Listing}

\newcommand{\incquery}{\textsc{EMF-IncQuery}}
\newcommand{\viatraemf}{\incquery}

\newcommand{\figref}[1]{Figure~\ref{#1}}

\newcommand{\secref}[1]{Section~\ref{#1}}

\newcommand{\appref}[1]{Appendix~\ref{#1}}

\definecolor{lvl2}{rgb}{0,0.5,0}
\definecolor{lvl1}{rgb}{0,0,0.5}
\definecolor{lvl0}{rgb}{0.5,0,0}

\makeatletter
\providecommand*{\toclevel@author}{0}
\providecommand*{\toclevel@title}{0}
\makeatother

\title{PN2SC Case Study: An \incquery{} solution\thanks{This work
was partially supported by the CERTIMOT (ERC\_HU-09-01-2010-0003),
 the T\'AMOP (4.2.2.B-10/1--2010-0009) projects.
 This research was realized in the frames of TÁMOP 4.2.4. A/1-11-1-2012-0001 
 ,,National Excellence Program – Elaborating and operating an inland student and researcher personal support system''. 
 The project was subsidized by the European Union and co-financed by the European Social Fund.}}
\author{
Benedek Izs\'o \and \'Abel Heged{\"u}s \and G\'abor Bergmann \and \'Akos Horv\'ath \and Istv\'an R\'ath
\institute{Budapest University of Technology and Economics,\\
Department of Measurement and Information Systems,\\
1117 Budapest, Magyar tud\'osok krt. 2.\\
\email{\{izso, hegedusa, bergmann, ahorvath, rath\}@mit.bme.hu}}
}
\def\titlerunning{PN2SC Case Study: An \incquery{} solution}
\def\authorrunning{Izs\'o, Heged{\"u}s, Bergmann, Horv\'ath, and R\'ath}

\begin{document}
\maketitle

\begin{abstract}
The paper presents a solution for the Petri-Net to Statecharts case study of the Transformation Tool Contest 2013, using \incquery{} and Xtend for implementing the model transformation.
\end{abstract}

\section{Introduction}
Automated model transformations are a key factor in modern model-driven system
engineering in order to query, derive and manipulate large, industrial models.
Since such transformations are frequently integrated to modeling environments,
they need to provide fast reaction time to support software engineers.

The objective of the \incquery{}~\cite{eiq-hompage} framework is to provide a declarative way to define queries over EMF models without needing to manually code imperative model traversals. \incquery{} extended the pattern language of Viatra (e.g.: with transitive closure, role navigation, match count) and tailored it to EMF models~\cite{iqpl}. The semantics of the pattern language is similar to VTCL (published previously), but the adaptation of the rule language is an ongoing work. \incquery{} uses the same incremental engine as Viatra, and latest developments extend this concept by providing a preliminary rule execution engine to perform transformations, however it is under heavy development, and the design of a dedicated rule language (instead of using the engine's API) is currently future work. The current case study aims at implementing the Petri-Net to Statecharts case study using \incquery{} as a rule engine. Conceptually, this new execution environment provides a mean to specify graph transformations (GT) as rules, where the LHS (left hand side) is defined with declarative \incquery{} graph patterns \cite{iqpl}, and the RHS (right hand side) as imperative model manipulations formulated in Xtend \cite{xtend}. Finally, the prototypical rule execution engine is configured from Java code, which automatically fire rules on match.

One case study of the 2013 Transformation Tool Contest describes a Petri-Net to Statecharts transformation \cite{GR13}. Main characteristics of the transformation are that it i) destructs the input (Petri-Net) model during the construction of the  output (Statechart) model, and ii) the transformation is divided into three phases: initialization, reduction and termination.

The rest of the paper is structured as follows: \secref{sec:overview} gives an overview of the implementation, \secref{sec:solution} describes the solution including design decisions, benchmark results and the solution for change propagation, and \secref{sec:conclusion} concludes our paper.

\section{Architecture overview}
\label{sec:overview}

\begin{figure}
        \centering
        \includegraphics[natwidth=11pt, natheight=11pt, width=.7\textwidth]{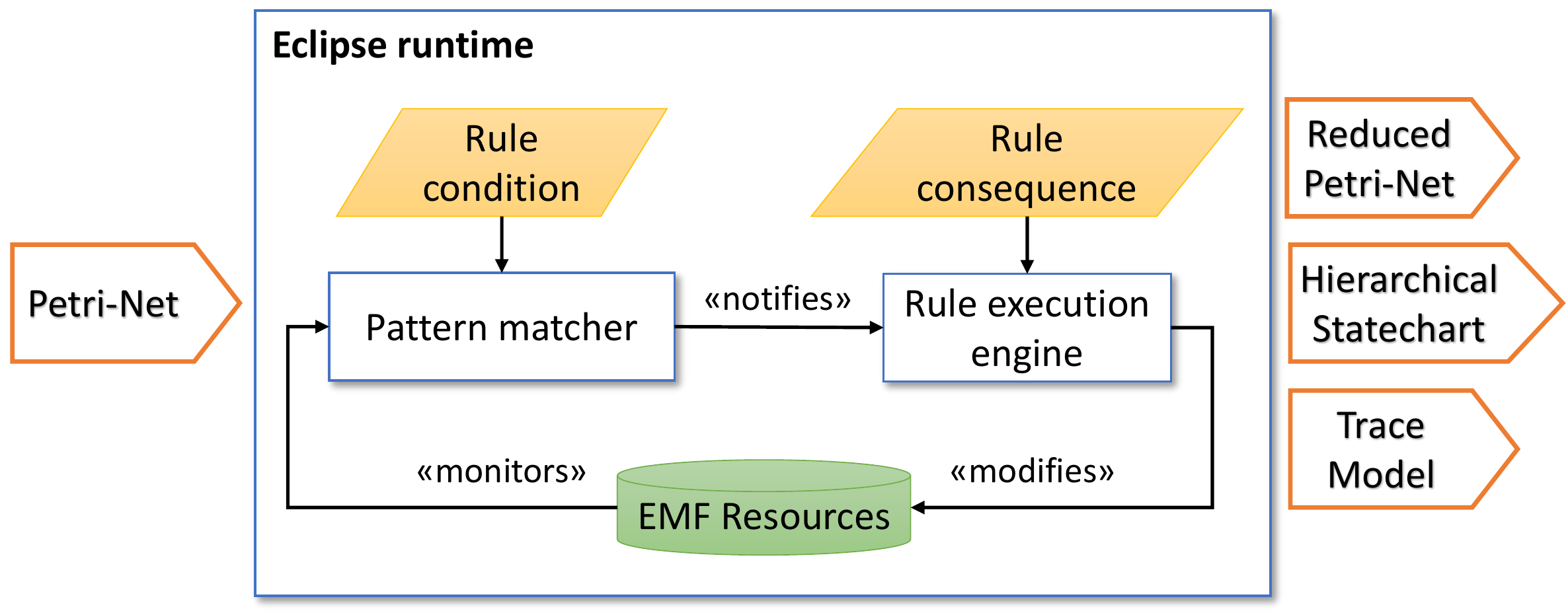}
        \caption{Overview of the specification and runtime}\label{fig:overview}
\end{figure}

The overview of the rule based solution is illustrated in \figref{fig:overview}. The input of the transformation is a Petri-Net, and as a result the reduced Petri-Net, a hierarchical statechart, and an auxiliary trace model is generated. The transformation is run in the Eclipse runtime: initially it reads the input Petri-Net resource, creates the output resources (organizing them into a resource set), then executes the transformation, and finally serializes the results into files. The transformation consists of three phases: the initial mapping, the Petri-Net reduction part (applying the OR and AND rules), and the termination phase (creation of the top Statechart elements). During the process, the \incquery{} incremental pattern matcher monitors the models for satisfiable rule conditions (from the rule set of the given phase), and on match notifies the rule execution engine. Based on the specified rule consequences, the rule engine modifies models of the resource set (reduces the Petri-Net and builds the Statechart), enabling new conditions to be satisfied, thus enabling new rules to fire. While there is some satisfied precondition, the engine fires them automatically.

% \begin{figure}
%         \centering
%         \begin{subfigure}[b]{0.48\textwidth}
%                 \centering
%                 \includegraphics[width=\textwidth]{../fig/overview_runtime.pdf}
%                 \caption{Runtime}
%                 \label{fig:runtime}
%         \end{subfigure}
%         ~ 
%         \begin{subfigure}[b]{0.35\textwidth}
%                 \centering
%                 \includegraphics[width=\textwidth]{../fig/overview.pdf}
%                 \caption{Specification}
%                 \label{fig:specification}
%         \end{subfigure}
%         \caption{Overview of the specification and runtime}\label{fig:overview}
% \end{figure}

The whole solution is written in three languages.  Rule condifions are formulated as \incquery{} graph patterns, while the rule consequences (model manipulations) in Xtend. These preconditions and rule actions are paired into rule specifications that are given to the execution engine using its Java based API.

\section{Solution}
\label{sec:solution}

\subsection{Specification}

The rule specification consists of two parts, which is illustrated in \figref{fig:ruleDefinition}. A partial solution of the AND rule demonstrates the formalization of its LHS and RHS.

The precondition of the AND rule is formulated in \incquery{} graph pattern language~\footnote{\incquery{} language: \url{http://wiki.eclipse.org/EMFIncQuery/UserDocumentation/QueryLanguage}}~\footnote{More examples and demos: \url{http://incquery.net/incquery/examples}}, as illustrated in \figref{fig:ruleCondition}. The pattern (named \emph{andPrecond}) can be satisfied in two ways (represented by two or-ed bodies), and returns satisfying \emph{Place-Transition} pairs, where the place \emph{P} is from the set of places from the precondition of the AND rule. The first case is described in lines 2-5, where transition \emph{T} has a \emph{pre}-place \emph{P} (line 2.), \emph{countPrePlaces} is the number of places with post-transition \emph{T} (line 3), which must be at least two (expressed by a check expression in line 4.). The \emph{T} post-transition must not have two pre-places with different pre- or post-transitions which is expressed by a negative application condition in line 5. The negatively called pattern finds two pre-places of \emph{T} (lines 13,17), and in the first case (lines 14-15) checks for a post-transition of \emph{P1} which is not a post-transition of \emph{P2}, while in the second case (lines 18-19) it checks for a pre-transition of P1 which is not a pre-transition of P2. The second case of the AND precondition (when the transition has at least two \emph{post}-places) can be formulated similarly. The whole code of the AND precondition and postcondition is described in \appref{app:andRule}.

\begin{figure}
        \centering
        \begin{subfigure}[b]{0.5\textwidth}
                \centering
                \listingIQPL{./fig/andPattern.vtcl}
                \caption{AND rule condition (\incquery{})}
                \label{fig:ruleCondition}
        \end{subfigure}%
        ~ \quad 
        \begin{subfigure}[b]{0.5\textwidth}
                \centering
                \listingXtend{./fig/andAction.xtend}
                \caption{AND rule processor (Xtend)}
                \label{fig:ruleProcessor}
        \end{subfigure} 
        
        %\begin{subfigure}[b]{0.37\textwidth}
        %       \centering
        %       \listingPseudo{../fig/andRuleDef.pseudo}
        %       \caption{AND rule specification}
        %       \label{fig:ruleDef}
        %\end{subfigure}
        \vskip -1.5ex
        \caption{Definition of the AND rule}\label{fig:ruleDefinition}
\end{figure}

The effect of the rule is achieved by executing imperative model editing commands, formulated in Xtend. Such model manipulations build up a processor, as illustrated in \figref{fig:ruleProcessor} for the AND rule. In lines 2-4 the set of places (\emph{placeSet}) is determined by checking whether the place is a pre-place of the transition, or a post-place. In lines 5-8 the new \emph{OR} and \emph{AND} states are created, connected, and put into the statechart model. Then mapped places (\emph{equiv(p)}) are moved below the newly created AND (lines 9-11). The place from the set of places selected by the pattern is reused, so after deleting old traces, a new trace is created for it, and other places are deleted from the Petri-Net (lines 12-15).

The specification of the AND rule binds the pattern \emph{andPrecond} as LHS, and \emph{andProcessor} as RHS using the Java API. These rules are executed automatically by the engine on match.

\subsection{Benchmark results}

The transformations were run on SHARE, on an Ubuntu 12.04, i686 architecture inside a VirtualBox. The CPU is an Intel Quad CPU Q9650 clocked at 3.00GHz, but in the virtualized environment only one is visible to the OS. The virtual computer has 1 GB of RAM, and 512 MB of swap space.

\begin{figure}[H]
  \centering
        \begin{subfigure}[b]{0.25\textwidth}
                \centering
                \includegraphics[natwidth=179pt,natheight=170pt, width=1\textwidth]{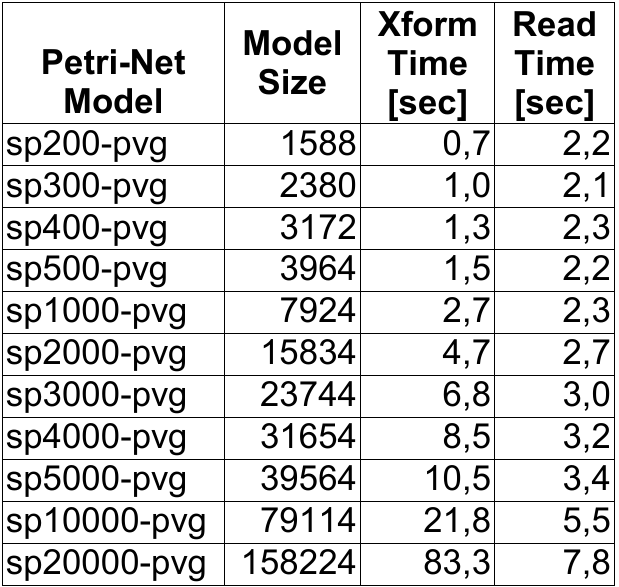}
                \caption{Benchmark results for EMF-IncQuery on SHARE (tabular)}
                \label{fig:resultsSHAREtable}
        \end{subfigure}
        ~
        \begin{subfigure}[b]{0.65\textwidth}
                \centering
                \includegraphics[natwidth=424pt,natheight=16pt, width=1\textwidth]{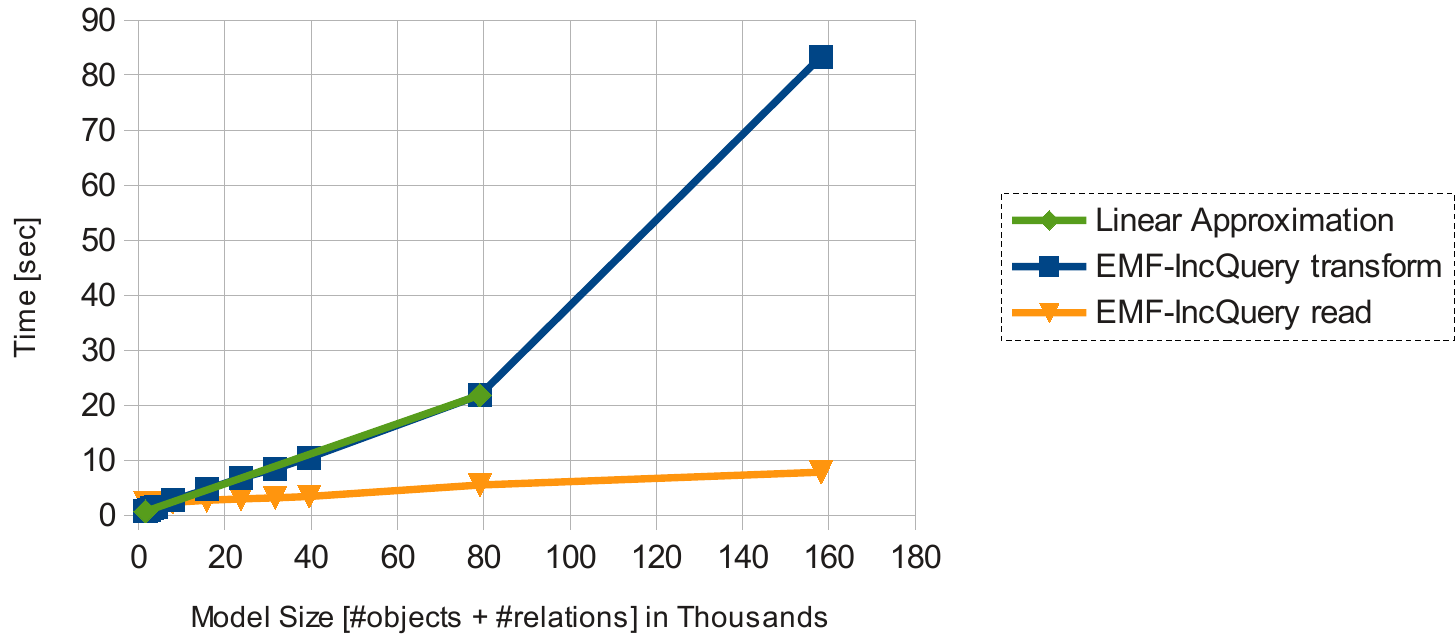}
                \caption{Benchmark results for EMF-IncQuery on SHARE (scatterplot)}
                \label{fig:resultsSHAREscatter}
        \end{subfigure}
        \caption{PN2SC benchmark results on SHARE for EMF-IncQuery}\label{fig:resultsSHARE}
\end{figure}

Results are displayed on \figref{fig:resultsSHARE}. \figref{fig:resultsSHAREtable} shows the numerical results in tabular form, where the first column is the name of the provided benchmark model, the second is the EMF model size, the third is the transformation time in seconds, and the fourth is the read time in seconds. The model size is the sum of all objects and relations of the EMF model.

\figref{fig:resultsSHAREscatter} displays the transformation time and model size on a scatterplot. It shows that \incquery{} scaled linearly up to 80 thousand elements (sp10000-pvg) (transforming the model in 22 secs), and ran for the model containing 158 thousand elements (sp20000-pvg). As the pattern matcher is a memory-intensive application, for the largest model more than 1 GB was necessary, which involved active swapping. This degraded runtime performance (obviously because hard disk is slower than RAM), and also because the CPU intensive kswapd and the transformation program shared the same CPU. Read times were not negligible, but were orders of magnitude less than the transformation time.
On one of our machines 10 GBs of memory could be given to the JVM, where transformations were run for models of all sizes. Here, the whole transformation for the sp20000-pvg model (largest model transformed on SHARE) was executed twice as fast as on SHARE. This effect can be probably attributed to less GC call, because for the smaller models,  runtimes were in orders of magnitude the same. Transforming the largest model on our machine took 87 minutes, however giving (and allocating) 15 GBs of memory instead of 10GBs, speed up the same transformation to 78 minutes.

These measured values are in accordance with the results published in the case study. The performance is linear for medium models, and exponential for large models, similarly to the GrGen.NET results. The hard (slow) parts for the \incquery{} tool was that this use case is model manipulation intensive, resulting in many intermediate changes of the result set of the patterns.

\subsection{Optimizations}
No test case specific optimizations were made, but for the whole system some special settings and best practices were applied. 
\emph{Finding common subpatterns} and extracting them into a pattern results in better performance (as the engine must process only once this part), and better maintainability (instead of copy-paste code). Such named pattern in Appendix \ref{app:andRule} is the \texttt{tranWithTwoPostPlaces} describing a structure that can be used in both (or-ed) bodies of \texttt{nonCommonTPre}. Named patterns can be called negatively (e.g. \texttt{postT}), and can be used as preconditions (e.g.: \texttt{andPrecond}).

% For moving EObject hierarchies between different containers, the \emph{engine provided cheapMoveTo()} function is used to avoid deletion and re-addition of many objects at engine level. \emph{Allocating the whole heap memory} that is available for the Java runtime at the beginning of a JVM execution turned to be useful, as resulted in less GC time. (On SHARE: ``-vmargs -Xmx896m -Xms896m''.)

\subsection{Transformation correctness and reproducibility}
The transformation runs correctly for the provided test cases on SHARE\footnote{http://is.ieis.tue.nl/staff/pvgorp/share/?page=ConfigureNewSession\&vdi=Ubuntu12LTS\_EIQ-PN2SC.vdi}, and the source code is also available on Github\footnote{https://github.com/izsob/TTC13-PN2SC-EIQ}. Automatic correctness validation was not implemented, but comparing the two models in the EMF tree editor shows equivalent structure. The transformation stops when multiple top level elements remain at the end, and creates a root element when only one top level element remains, enabling to inspect Statecharts with the provided GMF editors.

%The transformation can be reproduced, as the source code is available on Github~\cite{sourceOnGitHub}, and the solution can also be run on SHARE~\cite{demoOnSHARE}. In this virtualized environment the snapshot of the tool can be tried out, transformations can be run, and the source code is also available in Eclipse.

%A shortcut is created on the Desktop for each test case, as well as for running the benchmark transformations. Clicking on a shortcut starts the execution first by reverting the model to their initial state. Then executes the whole process, and writes the read time, transformation time and model save time in milliseconds to the console output. A shortcut to an Eclipse is also created which can be used to inspect the code. The host Eclipse contains the metamodels, while the ''Instances'' runtime Eclipse contains the transformation code.

\subsection{Change propagation}

As \incquery{} is an incremental technology, change propagation could be solved easily, using the same rule-based methodology. To handle the change of place, transition elements, or relations between them, patterns for precondition can be specified (matching only places, etc.), as illustrated in \ref{app:chpPatterns}.
 
Three rules are created to handle addition, deletion or name update of places and transitions with the processors described in lines \ref{cPProcessorBegin}-\ref{cPProcessorEnd} of \appref{app:chpProcessors}. The actual Petri-Net changes propagated to the target model are in lines \ref{cPManipBegin}-\ref{cPManipEnd} in \ref{app:chpProcessors}.

% \begin{figure}[H]
%         \centering
%         \begin{subfigure}[b]{0.28\textwidth}
%                 \centering
%                 \listingXtend{../fig/placeCP1.xtend}
%                 \caption{Place added}
%                 \label{fig:placeCPadd}
%         \end{subfigure}
%         ~ \quad 
%          \begin{subfigure}[b]{0.28\textwidth}
%                 \centering
%                 \listingXtend{../fig/placeCP2.xtend}
%                 \caption{Place deleted}
%                 \label{fig:placeCPdelete}
%         \end{subfigure}
%         ~ \quad 
%          \begin{subfigure}[b]{0.28\textwidth}
%                 \centering
%                 \listingXtend{../fig/placeCP3.xtend}
%                 \caption{Place name updated}
%                 \label{fig:placeCPupdate}
%         \end{subfigure}
%         
%         %\begin{subfigure}[b]{0.65\textwidth}
%         %        \centering
%         %        \listingPseudo{../fig/placeCPdef.pseudo}
%         %        \caption{Place rule definition}
%         %        \label{fig:placeCPdef}
%         %\end{subfigure}
%         \caption{Change propagation for handling Place manipulations.}\label{fig:placeCP}
% \end{figure}

This can be tested by running the ''PN2SC\_CP'' test case on SHARE from the runtime Eclipse. This performs changes on the transformed testcase1-in.petrinet. Snapshots of the changed Petri-Net and its propagations are saved in instances/snapshots, which can be inspected using the EMF tree editor.

\subsection{Tool support for debugging and refactoring}
As the transformation is written in three languages, debugging and refactoring is dependent on these languages, and on engine capabilities. Firings of the transformation can be debugged by placing breakpoints in the Xtend code, and debug messages of the execution engine can be turned on, which prints useful messages about rule firings and activations. Xtend and the \incquery{} pattern editors are based on Xtext, and while refactoring capabilities exist, they are sometimes limited. Debugging declarative \incquery{} graph patterns are impossible at runtime, but when snapshots are made from the model, the snapshot (EMF model) and queries can be loaded into the Query Explorer view, which is very handy to debug matches at a given point. The engine controller code can be debugged and refactored well, as it is written in Java. 

\section{Conclusion}
\label{sec:conclusion}

In this paper we have presented our \incquery{} based implementation for the Petri-Nets to Statechart case study. This is one of the the first cases where the prototipical execution engine based on \incquery{} is used as a rule engine, however, currently it has no dedicated rule language, and the engine is under heavy development.

The transformation is specified using declarative graph pattern queries over EMF models for rule preconditions, and Xtend code which can be executed to obtain the desired effect of the rule. Relying on incremental query evaluation of \incquery, the change propagations are also implemented.

%\nocite{*}
\bibliographystyle{eptcs}
\bibliography{bib/ttc13}

\begin{thebibliography}{1}
\providecommand{\bibitemdeclare}[2]{}
\providecommand{\surnamestart}{}
\providecommand{\surnameend}{}
\providecommand{\urlprefix}{Available at }
\providecommand{\url}[1]{\texttt{#1}}
\providecommand{\href}[2]{\texttt{#2}}
\providecommand{\urlalt}[2]{\href{#1}{#2}}
\providecommand{\doi}[1]{doi:\urlalt{http://dx.doi.org/#1}{#1}}
\providecommand{\bibinfo}[2]{#2}

\bibitemdeclare{inproceedings}{iqpl}
\bibitem{iqpl}
\bibinfo{author}{G{\'a}bor \surnamestart Bergmann\surnameend},
  \bibinfo{author}{Zolt{\'a}n \surnamestart Ujhelyi\surnameend},
  \bibinfo{author}{Istv{\'a}n \surnamestart R{\'a}th\surnameend} \&
  \bibinfo{author}{D{\'a}niel \surnamestart Varr{\'o}\surnameend}
  (\bibinfo{year}{2011}): \emph{\bibinfo{title}{{A Graph Query Language for EMF
  models}}}.
\newblock In: {\sl \bibinfo{booktitle}{Theory and Practice of Model
  Transformations, Fourth International Conference, ICMT 2011, Zurich}}, {\sl
  \bibinfo{series}{Lecture Notes in Computer Science}} \bibinfo{volume}{6707},
  \bibinfo{publisher}{Springer}, pp. \bibinfo{pages}{167--182},
  \doi{10.1007/978-3-642-21732-6\_12}.

\bibitemdeclare{misc}{xtend}
\bibitem{xtend}
\bibinfo{author}{\surnamestart {Eclipse.org}\surnameend}:
  \emph{\bibinfo{title}{{Xtend - Modernized Java}}}.
\newblock \bibinfo{howpublished}{\url{http://www.eclipse.org/xtend/}}.

\bibitemdeclare{misc}{eiq-hompage}
\bibitem{eiq-hompage}
\bibinfo{author}{\surnamestart Eclipse.org\surnameend} (\bibinfo{year}{2013}):
  \emph{\bibinfo{title}{{EMF-IncQuery}}}.
\newblock \bibinfo{howpublished}{\url{http://eclipse.org/incquery/}}.

\bibitemdeclare{inproceedings}{GR13}
\bibitem{GR13}
\bibinfo{author}{Pieter~Van \surnamestart Gorp\surnameend} \&
  \bibinfo{author}{Louis \surnamestart Rose\surnameend} (\bibinfo{year}{2013}):
  \emph{\bibinfo{title}{{The Petri-Nets to Statecharts Transformation Case}}}.
\newblock In \bibinfo{editor}{Louis~Rose \surnamestart Pieter
  Van~Gorp\surnameend} \& \bibinfo{editor}{Christian \surnamestart
  Krause\surnameend}, editors: {\sl \bibinfo{booktitle}{Sixth Transformation
  Tool Contest (TTC 2013)}}, {\sl \bibinfo{series}{EPTCS}}
  \bibinfo{volume}{this volume}.

\end{thebibliography}

\pagebreak
\appendix
\section{Appendix - PN2SC transformation code}
\label{app:xform}

\subsection{AND precondition as \viatraemf{} graph patterns}
\label{app:andRule}

The following code snippet shows the precondition of the AND rule with all dependent (called) patterns. Note that naming subpatterns (even simple ones) enhances performance, and these can be called (also negatively), or can be preconditions of rules.
\listingIQPL{./fig/andPrecond.eiq}

\newpage
The following Xtend code runs on the firing of the AND rule.
\listingXtend{./fig/andPostcond.xtend}

\subsection{Change propagation code}
\label{app:changeProp}

\subsubsection{Precondition patterns for the change-propagation task}
\label{app:chpPatterns}
\listingIQPL{./fig/ChangePropPatterns.eiq}

\subsubsection{Source model manipulation and target model modification functions in Xtend}
\label{app:chpProcessors}
\listingXtend{./fig/Pn2ScJobsChangeProp.xtend}

%\input{models10_appendix}
%\layout

\end{document}